\documentclass[aps,preprint,superscriptaddress]{revtex4}

\usepackage{graphicx}
\usepackage{bm}
\usepackage{amsmath}
\usepackage{amssymb}
\usepackage{hyperref}

\begin{document}

\title{Mass Density of Individual Cobalt Nanowires}

\author{L. Philippe}
\email{laetitia.philippe@empa.ch}
\affiliation{EMPA (Swiss Federal Laboratories for Materials Testing and Research), Feuerwerkerstrasse\,39, 3602\,Thun, Switzerland}

\author{B. Cousin}
\affiliation{EMPA (Swiss Federal Laboratories for Materials Testing and Research), Feuerwerkerstrasse\,39, 3602\,Thun, Switzerland}

\author{Z. Wang}
\email{zhao.wang@empa.ch}
\affiliation{EMPA (Swiss Federal Laboratories for Materials Testing and Research), Feuerwerkerstrasse\,39, 3602\,Thun, Switzerland}

\author{D.F. Zhang}
\affiliation{EMPA (Swiss Federal Laboratories for Materials Testing and Research), Feuerwerkerstrasse\,39, 3602\,Thun, Switzerland}

\author{J. Michler}
\affiliation{EMPA (Swiss Federal Laboratories for Materials Testing and Research), Feuerwerkerstrasse\,39, 3602\,Thun, Switzerland}

\begin{abstract}
The mass density of nanowires is determined using \textit{in-situ} resonance frequency experiments combined with quasi-static nanotensile tests. Our results reveal a mass density of 7.36 g/cm$^{3}$ on average which is below the theoretical density of bulk cobalt. Also the density of electrodeposited cobalt nanowires is found to decrease with the aspect ratio. The results are discussed in terms of the measurement accuracy and the microstructure of the nanowires.
\end{abstract}

\maketitle

During the last few years there has been increasing interest in the use of nanowires as key components in nanoelectronic devices \cite{Anantram2006}, due to their unique electric and mechanical properties. In particular, mechanical resonance of nanowires is of great interest for basic research \cite{Cleland1998,Martin1991} as well as for a wide range of applications in nanoelectromechanical systems such as sensors, actuators and field effect transistors \cite{Zook1992,Anantram2006,Craighead2000,Hoffmann2009,Friedli2007}. The fabrication of nanowires with low mass, tunable quality factor and natural frequency is crucial for the development of such nanodevices. The mass density of individual nanowires is one factor that determines resonance properties. It therefore plays a key role for potential applications of nanowires. 

However, very little work has been done to date studying the density of nanowires, due to the difficulties of mass detection at nanoscale. Densities at individual cross sections can be measured using high resolution microscopies \cite{Squier2001}, while it is almost impossible to measure the density along a whole wire using these techniques. Recently, resonance experiments have been used to study mechanical properties of free standing nanostructures \cite{Poncharal1999,Suryavanshi2004,Belov2008,Dikin2003,Utke2006,Hoffmann2006a}. These experiments are based on the Bernoulli-Euler theory, in which the natural resonant frequencies $f_{n}$ of a cantilevered beam can be written as:

\begin{equation}
\label{Eq:1}
f_{n}=\frac{(\beta_{n})^{2}}{2\pi} \sqrt{\frac{EI}{m L^{3}}}
\end{equation}

where $\beta_{n=1,2,3...} = 1.875, 4.694, 7.855 ...$ are constants for $n^{th}$ harmonic modes, $E$ is the axial elastic modulus, $I$ denotes the moment of inertia, $m$ is the mass and $L$ stands for the beam length. In a number of recent resonance experiments on nanowires, $E$ is calculated from this equation and $m$ is usually derived from the bulk density value. However, the density of nanowires could change with different fabrication procedures \cite{Hu1999,Dai2002}. In this study, we report on a solution to this problem for free-standing nanowires combining the resonance and the nanotensile experiments. We use \textit{in situ} resonant excitation of individual nanowires in a scanning electron microscope (SEM) to measure resonance frequencies, while the Young's modulus of nanowires is obtained from independent nanotensile experiments. The combination of these two experimental techniques allows us to calculate the mass density of individual nanowires using Eq. \ref{Eq:1}, unlike the techniques measuring only local density on individual cross sections using high resolution microscopies.

Cobalt nanowires are synthesized using electrochemical deposition in templates. The full experimental details of the process can be found elsewhere \cite{Philippe2007a}. The following electrolytes have been used for the cobalt solution: $CoSO_{4}$ (1M), $H_{3}BO_{3}$ (0.7M) and $NaCl$ (0.11M). The solution pH has been adjusted to $3.5$. Extraction of nanowires from the template has been performed by gold layer chemical dissolution ($I_{2}:2KI:10H_{2}O$), followed by dissoluting the polycarbonate membrane in dichloromethane.
 
To investigate the correlation between the density and mechanical properties of these electrodeposited cobalt nanowires, resonance experiments have been performed to measure the resonance frequency of free-standing samples. In the literature, vibrations of nanowires can be induced in one of their resonance modes by thermal vibration \cite{Treacy1996} or electric fields \cite{Poncharal1999}. In this work, vibrations of nanowires were induced by the oscillation of a piezoelectric actuator in a SEM chamber (Hitachi S-3600), in which samples were attached to a razor blade fixed on the piezoelectric actuator. Joints were made between the razor and nanowires by \textit{in-situ} carbon deposition using electron beam induced deposition. The amplitude-frequency curves were achieved by using secondary electron detection with a stationary beam near the sample. Peak of secondary electrons can be detected once the nanowires reach their maximal amplitude during the vibrations. In order to make these measurement, we increased the SEM magnification to 200,000 and adjusted the spot of the electron beam to make it slightly defocused. These techniques having more spatial interactions between the beam and the vibrating wires can increase the dynamic range \cite{Fujita2002,Ivobook2008}.

\begin{figure}[ht]
\centerline{\includegraphics[width=14cm]{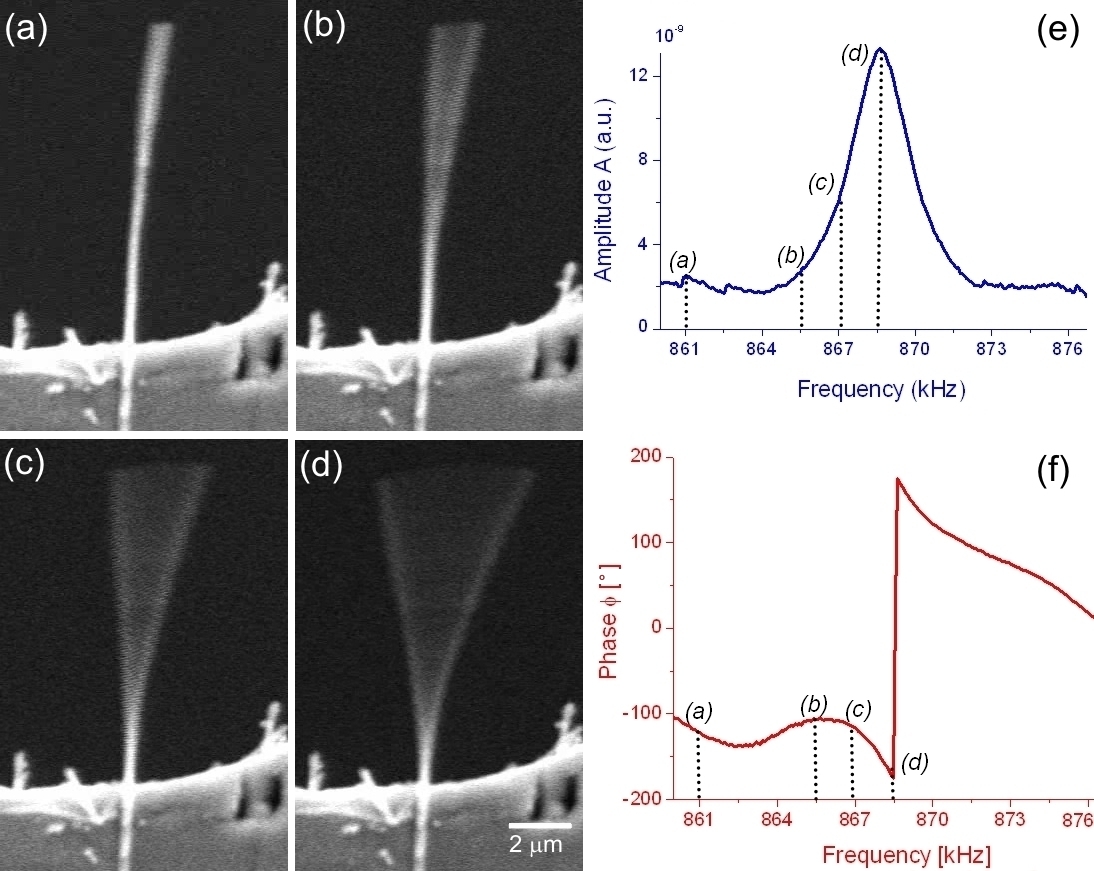}}
\caption{\label{fig:2}
(Color online) (a)-(d): SEM images of the resonance of a free-standing cobalt nanowire ($L \approx 15\mu$m) at fundamental mode. Vibrations were induced by piezo-mechanical excitation and detected by measuring the secondary electron signal created by electric interactions between the wire and the electron beam. (e)-(f): a wide spectrum of the amplitude response which locates in the peak in the vicinity of $868$ KHz. A real-time video-recording of this experiment is available online \cite{movie1}.
}
\end{figure}

An example of our resonance experiments are shown in Figure \ref{fig:2}, in which we can see different stages of the vibration of a nanowire (a-d). In Figure \ref{fig:2} (e-f), an overview spectrum was acquired by means of the stationary beam technique, in which we locate resonance peaks sweeping the excitation frequency through the full available detection bandwidth up to $1.2$ MHz and measuring amplitude and phase response. We can see that the wire vibrate at the fundamental harmonic mode and the maximum amplitude of deflection can be measured from SEM images.

To measure the axial elastic modulus of these nanowires, quasi-static tensile tests have been performed using a microelectromechanical system (MEMS) based tensile testing stage, which consists of a comb drive actuator, a force sensor and a gap in between. An individual cobalt nanowire was bridged on the gap by pick-and-place nanomanipulation performed inside a SEM and fixed at the two ends via focused electron/ion beam induced deposition \cite{Utke2008}. A uniaxial voltage-controlled tensile force was applied to the specimen through the electrostatic actuator. The tensile load was measured from the displacement of the force sensor with a known spring constant. Series of high-magnification SEM images were taken during the tensile experiments, from which the deformation of the specimen elongation could be extracted by image analysis with a homemade program based on a cross correction algorithm \cite{Hoffmann2007}. Further details on the instrument and the experiments can be found in Ref. \cite{Zhang2010,Zhang200902}.  

\begin{figure}[ht]
\centerline{\includegraphics[width=14cm]{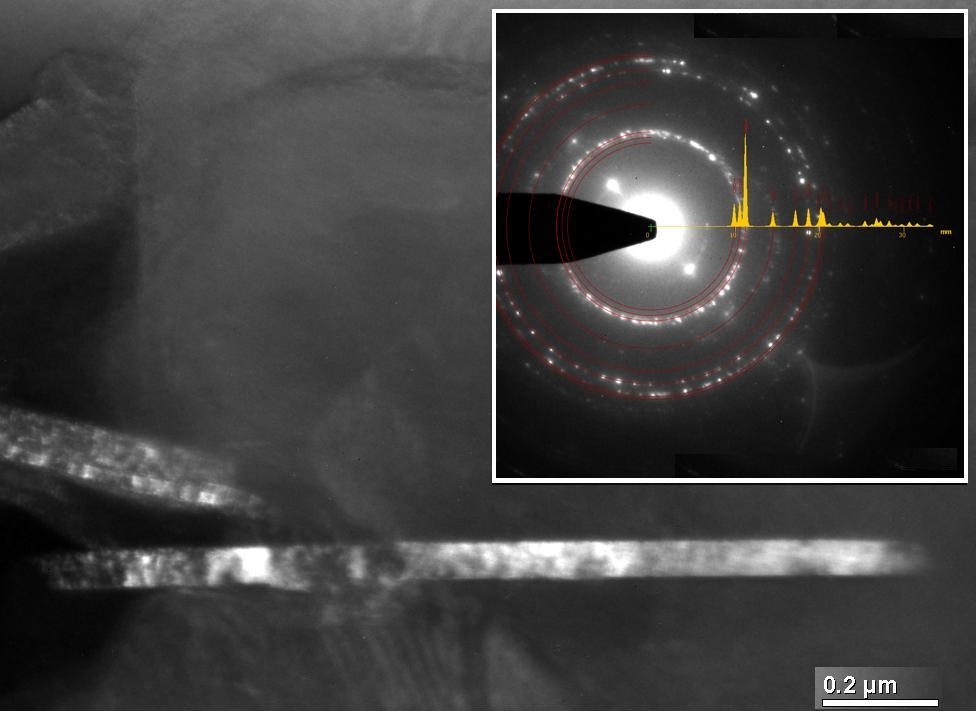}}
\caption{\label{fig:0} TEM image along corresponding diffraction pattern of a cobalt nanowire. Inset: pattern reveals that this cobalt nanowire exhibits variations of crystal orientations (polycrystalline). The plot on the pattern shows the intensity of the diffracted beams with respect to the diffraction angle. 
}
\end{figure}

The nanostructures of electrodeposited cobalt nanowires were analyzed by using a transmission electron microscope (TEM). It has been found that these wires exhibit crystal orientation variations along their axis (see Fig. \ref{fig:0}), and that the grain sizes vary roughly from $10$ to $150$ nm. These local variations can be due to variations of metal concentration and pH value in solution during the synthesis processes, which can reduce reaction kinetics during the deposition \cite{Philippe2007a, Philippe2008}. From a poromechanical point of view, these density variations can be expected to influence significantly the elastic modulus \cite{Choi2007,Varghese2008} and the quality factor \cite{Smith2008}. 

\begin{figure}[ht]
\centerline{\includegraphics[width=14cm]{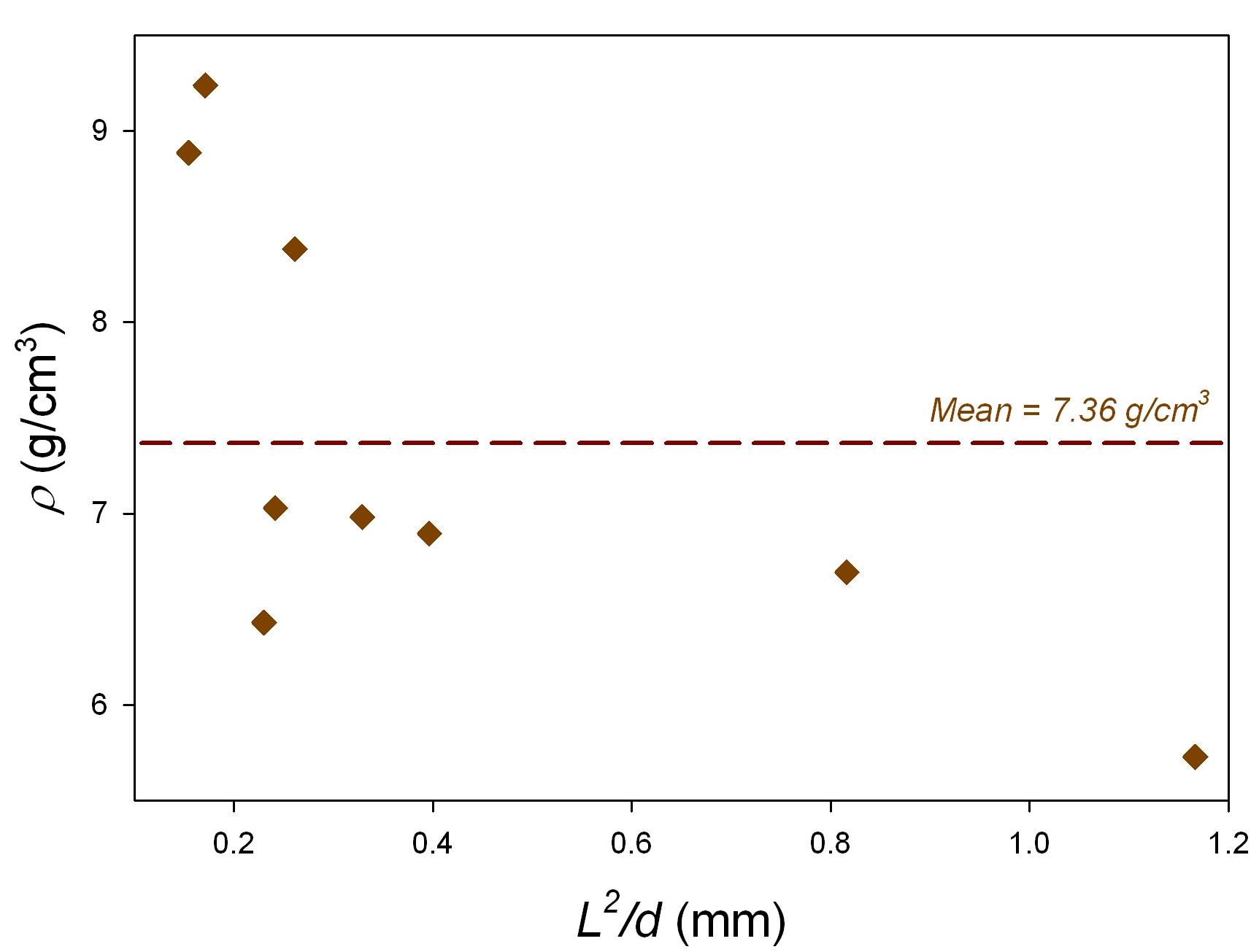}}
\caption{\label{fig:3}
(Color online) Densities of electrodeposited cobalt nanowires calculated from Eq. \ref{Eq:9}. Dashed line shows the average value.}
\end{figure}

The quality factor $Q$ quantifies the energy dissipation in a vibration structure to the environment. In our case of high vacuum condition, $Q$ mainly describes the dissipation to the supports. Here quality factors for the fundamental resonance mode of cobalt nanowires in high vacuum at room temperature were determined from the slope of the phase curve at resonance using the following equation:

\begin{equation}
\label{Eq:Q}
Q= \frac{f_{0}}{2} \left| \frac{d \phi_{(f_{0})} } {df} \right|
\end{equation}

where $d \phi / df$ is the phase variation \cite{Friedli2009,Perisanu2007}. The natural frequencies $f_{0}$ of the nanowires have been measured in resonance experiments, their values are plotted in Figure \ref{fig:3}.

Results of $Q$ (see Table \ref{table1}) based on frequencies obtained from resonance experiments show that $Q$ roughly increases with increasing volume of nanowires $V=0.25\pi Ld^{2}$. This is in good agreement with the conclusion in Ref. \cite{Perisanu2007}. However, a good reproducibility of the quality factor measured for a given volume, as well as an expected increase of $Q$ with $V$ was observed in both sets of measurements. These results show the possibility to make nanowires with tunable quality factor by controlling the wire size. This is an important issue for applications of nanowires in nanoelectronics.

\begin{table}[h]
\caption{\label{table1} $Q$ factors of cobalt nanowires}
\begin{center}
\begin{tabular}{ccc|ccc}
\hline \hline
   sample & $V$ $(\mu m^{3})$ & Q &   sample   & $V$ $(\mu m^{3})$ & Q \\
\hline 
1 &1.57      & 589  & 4&0.42      & 338   \\
2 &0.92      & 581   & 5&0.29      & 328       \\
3 &0.57     & 444   & 6&0.22      & 324.5      \\
\hline \hline
\end{tabular}
\end{center}
\end{table}

Eight cobalt nanowires were tested in the nanotensile experiments (see Ref. \cite{Zhang2010}). The measured average value of the Young's modulus is $75.28 \pm 14.6$ GPa, which is much lower than that of the bulk ($209$ GPa). One probable reason for the reduction in the Young's modulus is the presence of defects (e.g. pores and grain boundaries) in the nanowires. The defect-induced mechanical softening effects were previously reported for both Nanostructures and bulk materials \cite{Sakai1998,Kovacik1999,Varghese2008}.   

As shown in Eq. \ref{Eq:1}, the resonance frequency depends on the elastic modulus and the density of nanowires. If the cross section is considered as a cylinder $I = \pi d^{4} /64$, the frequency $f_{0}$ in fundamental mode can be written as a function of wire density $\rho$ as follows:

\begin{equation}
\label{Eq:9}
f_{0}=\frac{ d \beta_{0}^{2} }{8 \pi L^{2}} \sqrt{\frac{E}{\rho}}
\end{equation}

where $d$ and $L$ are the diameter and the length of nanowires, respectively. Putting the values of $f_{0}$ from resonance experiments and those of Young's modulus from tensile experiments into Eq. \ref{Eq:9}, we can calculate the average density of each wire (see results plotted in Fig. \ref{fig:3}). We can see that the densities decrease with increasing ratio $L^{2}/d$. The average density value of our electrodeposited cobalt nanowires is about $7.36$ $g/cm^{3}$, which is about $83\%$ of the bulk value ($8.9$ $g/cm^{3}$). A value lower than the theoretical density is expected from the results of the TEM analysis which shows a nanocrystalline structure and the low elastic modulus of the nanowires. From bulk material for instance it is known that a pore content of 17\% volumn would already decrease the bulk modulus by about $50$\% \cite{Kovacik1999,Kovacik2001}. Following our error estimation approach taking into account the imprecision in the measurement of the dimensions, the estimated maximum error in the density is approximately $\pm 75$\%, including $2\times 5$\% from the diameter, $4\times 10$\% from the length, $5$\% from the frequency and $20$\% from the Young's modulus \cite{Zhang2010}. Note that the errors from the attachment is difficult to estimate since the substrate is not totally planar and the clamping depth of the nanowire into the substrate is difficulte to be estimated.

In conclusion, we have measured the density of electrodeposited cobalt nanowires, combining nanoresonance and nanotensile experiments. This combination allows calculating the mass density of individual nanowires from the experimentally measured resonance frequency and elastic modulus, using a classical formula of elasticity theory. It is found that the mass density of nanowires decreases with the aspect ratio $L^{2}/d$ and its average value is determined to be about $7.36$ $g/cm^{3}$. We also found that the average resonance frequency and Young's modulus of these nanowires are about $0.94$ MHz and $75.28 \pm 14.6$ GPa, respectively. The low values of the density and the elastic modulus correspond to the nanostructure revealed by a TEM analysis. Moreover, the quality factor of these nanowires is found to increase with the nanowire volume. This reveals a possibility to fabricate nanowires with well-defined quality factors. These results are expected to be useful in particular for the potential applications of nanowires in nanoelectronic devices or nanoelectromechanical systems.  

\section*{Acknowledgement}
Financial support by the Swiss State Secretariat for Education and Research FP7-NMP HYDROMEL (under contract No. 026622-2) is gratefully acknowledged. We thank P. Stadelmann at EPFL for sharing the TEM pictures and E. Balic, I. Utke and V. Friedli for useful discussion. We note that L.P. and B.C. contributed equally to this work.

\section*{Figures}

\newpage

\newpage

\end{document}